\documentclass[letterpaper, 10 pt, conference]{ieeeconf}  

\usepackage{graphicx}      
\usepackage{cite}        
\usepackage{siunitx}
\usepackage{amsmath,cancel}
\usepackage{array}
\usepackage{amsmath}
\usepackage{tikz}
\usepackage{mathdots}
\usepackage{yhmath}
\usepackage{cancel}
\usepackage{color}
\usepackage{multirow}
\usepackage{amssymb}
\usepackage{gensymb}
\usepackage{tabularx}
\usepackage{booktabs}
\usepackage{arydshln}
\newtheorem{rem}{Remark}
\newtheorem{thm}{Theorem}
\usepackage[font=footnotesize,labelfont=bf]{caption}
\usepackage[font=footnotesize,labelfont=bf]{subcaption}
\IEEEoverridecommandlockouts                              

\title{\LARGE \bf
Stacking Integrators Without Sacrificing the Overshoot in Reset Control Systems 
}

\author{Nima Karbasizadeh$^{1}$ and S. Hassan HosseinNia$^{1}$
\thanks{This work was supported by NWO, through OTP TTW project \#16335.}
\thanks{$^{1}$Department of Precision and Microsystem Engineering, Delft University of Technology, Delft, The Netherlands,
        }
\thanks{{\tt\small N.KarbasizadehEsfahani@tudelft.nl}}
\thanks{{\tt\small S.H.HosseinNiaKani@tudelft.nl}}%
}

\begin{document}

\maketitle
\thispagestyle{empty}
\pagestyle{empty}

\begin{abstract}

According to the well-known loop-shaping control design approach, the steady-state precision of control systems can be improved by stacking integrators. However, due to the waterbed effect in linear control systems, such an action will worsen the transient response by increasing overshoot and creating wind-up problems. This paper presents a new architecture for rest control systems that can significantly decrease the overshoot and create a no-overshoot performance even in presence of stacked integrators. The steady-state analysis of the proposed system will also show that improved precision expected due to stacked integrators can be achieved as well. A numerical simulation study is presented to verify the results and the tuning guide presented. 

\end{abstract}

\section{INTRODUCTION}

Increase the gain in lower frequencies as much as possible, this is one of the basic controller design criteria, particularly in precision motion control. The criterion comes from the well-known loop shaping technique for designing controllers in frequency domain~\cite{schmidt2020design}.  Such a design allows for reducing the steady-sate error due to reference tracking and disturbances. One of the possible ways to increase the gain at lower frequencies is to stack integrators. However, a well-known fundamental limitation of linear control systems precludes stacking too many integrators, the so-called ``Waterbed Effect''~\cite{bode1945network}. One can interpret the waterbed effect for stacking integrators as excessive overshoot in transient response. However nonlinear control systems are not inherent to this limitations.\\
Reset control is one of the nonlinear control techniques which introduces a simple non-linearity to control systems but allows the the control system to overcome the linear control limitations. Reset controllers were first proposed by Clegg~\cite{clegg1958nonlinear} in forms of a reset integrator which can improve the transient response of a control system.  In order to address the drawbacks and exploiting the benefits, the idea was later extended to more sophisticated elements such as ``First-Order Reset Element''~\cite{horowitz1975non,krishnan1974synthesis} and ``Second-Order Reset Element''~\cite{hazeleger2016second} or using Clegg's integrator in form of PI+CI~\cite{banos2007definition} or resetting the state to a fraction of its current value, known as partial resetting~\cite{beker2004fundamental}. Reset control has also recently been used to approximate the complex-order filters~\cite{valerio2019reset,saikumar2019constant}. Advantage of using reset control over linear control has been shown in many studies especially in precision motion control ~\cite{BISOFFI202037,banos2011reset,karbasizadeh2020benefiting,beker2004fundamental,dastjerdi2021frequency,wu2006reset,wu2006reset,zheng2000experimental,chen2020development,chen2001analysis,karbasizadeh2021fractional}. However, these studies are mostly focused on solving one problem. For example they either improve transient or steady-state response of the system while paying little or no attention to the other.\\
One of the recent studies introduces a new reset element called ``Constant-in-Gain, Lead-in-Phase'' (CgLp) element which is proposed based on the loop-shaping concept~\cite{saikumar2019constant}. Describing Function (DF) analysis of this element shows that it can provide broadband phase lead while maintaining a constant gain. Such an element is used in the literature to replace some part of the differentiation action in PID controllers as it will help improve the precision of the system according to loop-shaping concept~\cite{karbasizadeh2021fractional,karbasizadeh2020benefiting,dastjerdi2021frequency,saikumar2019constant}.\\
In~\cite{karbasizadeh2020benefiting,karbasizadeh2021fractional}, it is suggested that DF analysis for reset control systems can be inaccurate as it neglects the higher-order harmonics created in response of reset control systems. It is concluded that higher-order harmonics can adversely affect the steady-state precision of the system. \\
One of the benefits of providing phase lead through CgLp is improving the transient response properties of the system, as it is shown that it reduces the overshoot and settling time of the system. However, the way to achieve this goal is not only through phase compensation around cross-over frequency. It is shown in~\cite{cai2020optimal} that since reset control systems are nonlinear systems, the sequence of elements in control loop affects the output of the system. It was shown that when the lead elements are placed before reset element, it can improve the overshoot of the system. However, no systematic approach is proposed there for further improving the transient response. In~\cite{ZHAO201927}, it is shown that by changing the resetting condition of reset element to reset based in its input and its derivative, overshoot limitation in linear control, systems can be overcome. This limitation has also been broken using the same technique in another hybrid control system called ``Hybrid Integrator Gain System" (HIGS)~\cite{vanden2020hybrid}. However, in these studies the effect of such an action on steady-state performance of the system is not addressed.  \\
The main contribution of this paper is to propose a new architecture for CgLp element which can benefit from increased steady-state precision due to stacked integrators while compensating the increased overshoot and even eliminating the overshoot. This paper shows that this architecture will not even increase the maximum of control input and thus is less prone to wind-up problems. A guideline for tuning the proposed architecture is also provided.\\
The remainder of this paper is organized as follows:\\
Section~\ref{sec:prim} introduces the preliminaries of the research. Section~\ref{sec:CR} will present the proposed reset control architecture and its properties. Section~\ref{sec:loop} will introduce the suggested control loop design and Section~\ref{sec:transient} will present a numerical study on closed-loop transient response of the proposed control system and provides the tuning guideline. Section~\ref{sec:steady} presents the closed-loop steady-state performance analysis of the proposed control system. At last, paper closes with conclusion and ongoing work tips.
\section{Preliminaries}
\label{sec:prim}
This section will discuss the preliminaries of this study.
\subsection{General Reset Controller}
The general form of reset controllers used in this study is as following:
\begin{align}
	\label{eq:reset}
	{{\sum }_{R}}:=\left\{ \begin{aligned}
		& {{{\dot{x}}}_{r}}(t)={{A_r}}{{x}_{r}}(t)+{{B_r}}e(t),&\text{if }e(t)\ne 0\\ 
		& {{x}_{r}}({{t}^{+}})={{A}_{\rho }}{{x}_{r}}(t),&\text{if }e(t)=0 \\ 
		& u(t)={{C_r}}{{x}_{r}}(t)+{{D_r}}e(t) \\ 
	\end{aligned} \right.
\end{align}
where $A_r,B_r,C_r,D_r$ denote the state space matrices of the base linear system (BLS) and reset matrix is denoted by $A_\rho=\text{diag}(\gamma_1,...,\gamma_n)$ which contains the reset coefficients for each state. $e(t)$ and $ u(t) $ represent the input and output for the reset controller, respectively.
\subsection{$H_\beta$ condition}
The quadratic stability of the closed loop reset system when the base linear system is stable can be examined by the following condition~\cite{beker2004fundamental,Guo:2015}.
\begin{thm}             
	There exists a constant $\beta \in \Re^{n_r\times 1}$ and positive definite matrix $P_\rho \in \Re^{n_r\times n_r}$, such that the restricted Lyapunov equation
	\begin{eqnarray}
		P > 0,\ &A_{cl}^TP + PA_{cl} &< 0\\
		&B_0^TP &= C_0
	\end{eqnarray}
	has a solution for $P$, where $C_0$ and $B_0$ are defined by
	\begin{align}
		C_0=\left[\begin{array}{ccc}
			\beta C_{p} & 0_{n_r \times n_{nr}} & P_\rho
		\end{array}\right] , & &  B_0=\left[\begin{array}{c}
			0_{n_{p} \times n_{r}}\\
			0_{n_{nr} \times n_{r}}\\
			I_{n_r}
		\end{array}\right].
	\end{align}
	and 
	\begin{equation}
		A_{\rho}^TP_\rho A_{\rho} - P_{\rho} \le 0
	\end{equation}
	where $A_{cl}$ is the closed loop A-matrix including an LTI plant dynamics.
	$n_r$ is the number of states being reset and $n_{nr}$ being the number of non-resetting states and  ${n_{p}}$ is the number states for the plant.
	$A_p,B_p,C_p,D_p$ are the state space matrices of the plant.
\end{thm}
In~\cite{dastjerdi2020frequency}, the $H_\beta$ condition is extended to systems where reset element is not the first element in the loop, in other words, the input to the reset element is a shaped error signal. The stability analysis of elements presented in this paper can be done using theories in~\cite{dastjerdi2020frequency}.  
\subsection{Describing Functions}
Describing function analysis is the known approach in literature for approximation of frequency response of nonlinear systems like reset controllers\mbox{\cite{guo2009frequency}}. However, the DF method only takes the first harmonic of Fourier series decomposition of the output into account and neglects the effects of the higher order harmonics. This simplification can be significantly inaccurate under certain circumstances~\cite{karbasizadeh2020benefiting}. The ``Higher Order Sinusoidal Input Describing Function'' (HOSIDF) method has been introduced in\mbox{\cite{nuij2006higher}} to provide more accurate  information about the frequency response of nonlinear systems by investigation of higher-order harmonics of the Fourier series decomposition. In other words, in this method, the nonlinear element will be replaced by a virtual harmonic generator. This method was developed in\mbox{\cite{saikumar2021loop,dastjerdi2020closed}} for reset elements defined by Eq.~({\ref{eq:reset}}) as follows:
\begin{align}  \nonumber
	\label{eq:hosidf}
	& H_n(\omega)=\left\{ \begin{aligned}
		& C_r{{(j\omega I-A_r)}^{-1}}(I+j{{\Theta }}(\omega ))B_r+D_r,~~ n=1\\ 
		& C_r{{(j\omega nI-A_r)}^{-1}}j{{\Theta }}(\omega )B_r,~~~\qquad\text{odd }n> 2\\ 
		& 0,~\quad\qquad\qquad\qquad\qquad\qquad\qquad~\text{even }n\ge 2\\ 
	\end{aligned} \right. \\
	&\begin{aligned}
		& {{\Theta }}(\omega )=-\frac{2{{\omega }^{2}}}{\pi }\Delta (\omega )[{{\Gamma }}(\omega )-{{\Lambda }^{-1}}(\omega )] \\  
		& \Lambda (\omega )={{\omega }^{2}}I+{{A_r}^{2}} \\  
		& \Delta (\omega )=I+{{e}^{\frac{\pi }{\omega }A_r}} \\  
		& {{\Delta }_{\rho}}(\omega )=I+{{A}_{\rho}}{{e}^{\frac{\pi }{\omega }A_r}} \\  
		& {{\Gamma }}(\omega )={\Delta }_{\rho}^{-1}(\omega ){{A}_{\rho}}\Delta (\omega ){{\Lambda }^{-1}}(\omega ) \\
	\end{aligned} 
\end{align}
where $H_n(\omega)$ is the $n^{\text{th}}$ harmonic describing function for sinusoidal input with frequency of $\omega$. \\
\begin{figure}[t!]
	\centering
	\includegraphics[width=\columnwidth]{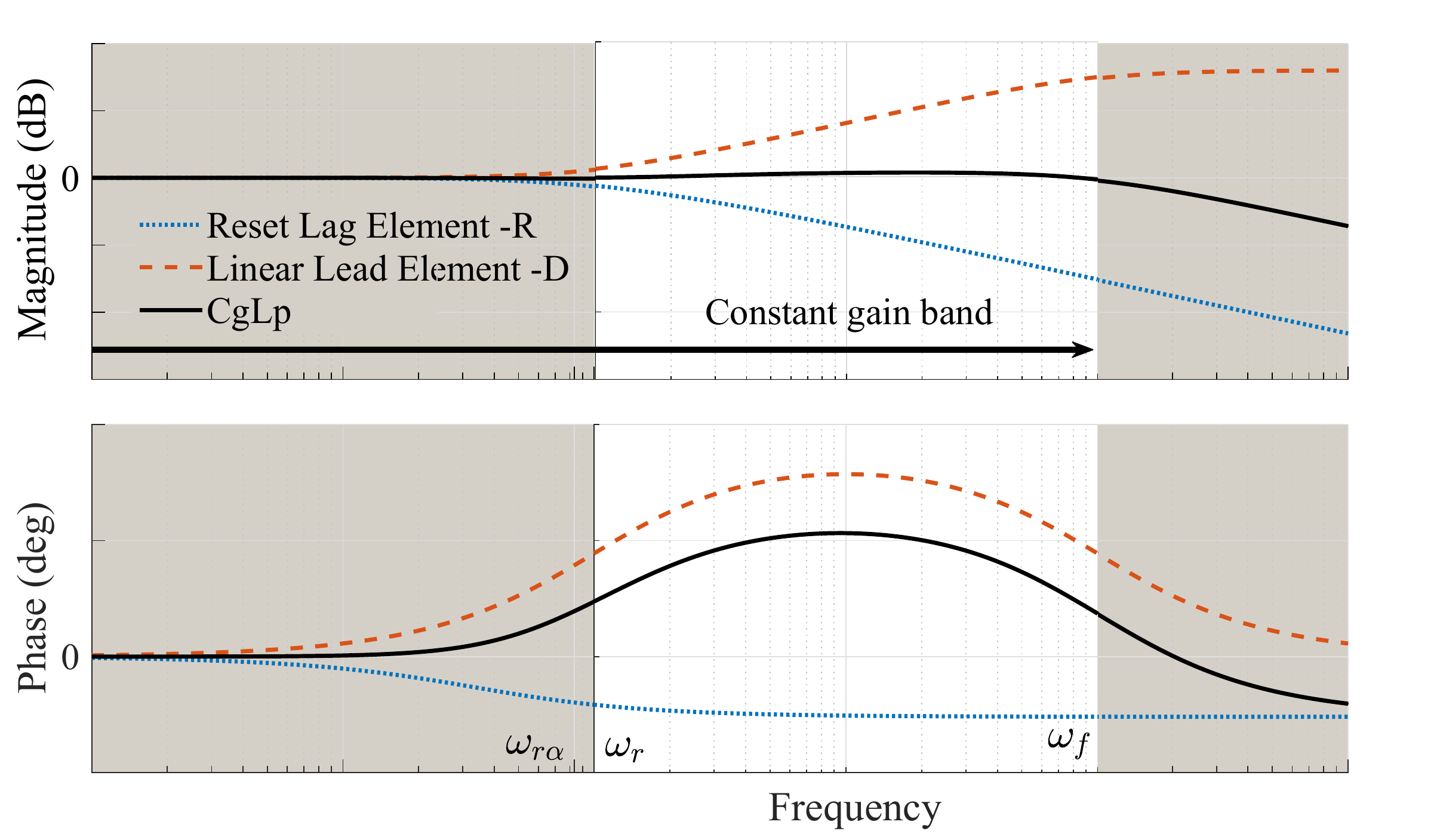}
	\caption{The concept of using combination of a reset lag and a linear lead element to form a CgLp element. The figure is  from~\cite{saikumar2019constant}.}
	\label{fig:cglp}
\end{figure}
\subsection{CgLp}
CgLp is a broadband phase compensation reset element which has a first harmonic constant gain behaviour while providing a phase lead~\cite{saikumar2019constant}. This element consists in a reset lag element in series with a linear lead filter, namely ${\sum}_R$ and $D$. For FORE CgLp:
\begin{align}
	\label{eq:fore}
	&{\sum}_R=\cancelto{\gamma}{\frac{1}{{s}/{{{\omega }_{r }}+1}\;}},&D(s)=\frac{{s}/{{{\omega }_{r\alpha}}}\;+1}{{s}/{{{\omega }_{f}}}\;+1}
\end{align}
where $\omega_{r\alpha}=\alpha \omega_r$, $\alpha$ is a tuning parameter accounting for a shift in corner frequency of the filter due to resetting action,  and $[\omega_{r},\omega_{f}]$ is the frequency range where the CgLp will provide the required phase lead. The arrow indicates the resetting action as described in Eq.~\eqref{eq:reset}, i.e., the state matrix of the element is multiplied by $A_\rho$ when the reset condition is met. \\
CgLp provides the phase lead by using the reduced phase lag of reset lag element in combination with a corresponding lead element to create broadband phase lead. Ideally, the gain of the reset lag element should be canceled out by the gain of the corresponding linear lead element, which creates a constant gain behavior. The concept is depicted in Fig.~\ref{fig:cglp}.\\
\section{Proposed Architecture for Continuous Reset (CR) Elements}
\label{sec:CR}
\begin{figure}[t!]
	\vspace{5pt}
	\centering
	\resizebox{\columnwidth}{!}{

		\tikzset{every picture/.style={line width=0.75pt}} 
		
		\begin{tikzpicture}[x=0.75pt,y=0.75pt,yscale=-1,xscale=1]
			
			\draw  [line width=0.75]  (63,40) -- (133,40) -- (133,92.07) -- (63,92.07) -- cycle ;
			\draw  [line width=0.75]  (187.5,43.17) -- (260,43.17) -- (260,92) -- (187.5,92) -- cycle ;
			\draw [line width=0.75]    (182.5,104.08) -- (249.53,27.34) ;
			\draw [shift={(251.5,25.08)}, rotate = 491.13] [fill={rgb, 255:red, 0; green, 0; blue, 0 }  ][line width=0.08]  [draw opacity=0] (10.72,-5.15) -- (0,0) -- (10.72,5.15) -- (7.12,0) -- cycle    ;
			\draw [line width=0.75]    (0,71) -- (63,71) ;
			\draw [line width=0.75]    (133,70.75) -- (188,71) ;
			\draw  [line width=0.75]  (320,42.17) -- (389,42.17) -- (389,91) -- (320,91) -- cycle ;
			\draw [line width=0.75]    (260,70) -- (320,70) ;
			\draw [line width=0.75]    (389.5,70) -- (436.5,70) ;
			\draw  [dash pattern={on 4.5pt off 4.5pt}]  (160.5,70.87) -- (160,81) ;
			\draw  [dash pattern={on 4.5pt off 4.5pt}]  (160,81) -- (187,81) ;
			
			\draw (415.5,54.5) node  [font=\Large]  {$u( t)$};
			\draw (31,55.5) node  [font=\Large]  {$e( t)$};
			\draw (257,3.4) node [anchor=north west][inner sep=0.75pt]  [font=\Large]  {$A_{\rho }$};
			\draw (80,54.15) node [anchor=north west][inner sep=0.75pt]  [font=\LARGE]  {$L( s)$};
			\draw (201,52.4) node [anchor=north west][inner sep=0.75pt]  [font=\LARGE]  {$\sum{}_{R}$};
			\draw (335,52.4) node [anchor=north west][inner sep=0.75pt]  [font=\LARGE]  {$R( s)$};
			\draw (160,55.5) node  [font=\Large]  {$x_{1}( t)$};
			\draw (294,56) node  [font=\Large]  {$x_{2}( t)$};

		\end{tikzpicture}
	}
	\caption{Proposed architecture for reset elements which includes a lead element, $L(s)$ before the reset element and ints inverse after the reset element. The proposed lead is $L(s)=\frac{s/\omega_l+1}{s/\omega_h+1}$ and $R(s)=\frac{1}{s/\omega_l+1}$.}
	\label{fig:new_arch}
\end{figure}
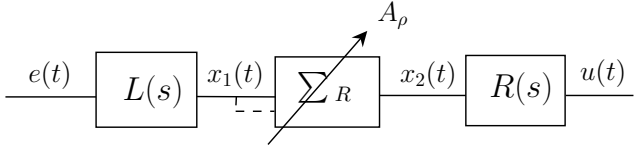
The new architecture which this paper proposes consists of adding a first-order lead element, $L(s)$, before the reset element and adding the inverse of it, which is basically a lag element, after the reset element. Fig.~\ref{fig:new_arch} depicts the new architecture in which
\begin{equation}
	L(s)=\frac{s/\omega_l+1}{s/\omega_h+1},\quad R(s)=\frac{1}{s/\omega_l+1}.
	\label{eq:L}
\end{equation}
The presence of the denumerator in $L(s)$ is to make transfer function proper. If $\omega_h$ is large enough, $R(s)\approx L^{-1}(s)$ in low frequencies. In the context of linear control systems, adding these two elements would almost have no effect on the output of the system, provided the internal states stability. However, in the context of nonlinear control systems, the output of the system will be changed significantly. \\
In this new architecture the resetting condition is changed from $e(t)=0$ to $x_1(t)=0$. Considering that $\omega_h$ is large enough, the new resetting condition can be approximated as  
\begin{equation}\label{eq:reset_law}
	x_1(t)={\dot{e}(t)}/\omega_l+e(t)=0.
\end{equation}
The new reset element resets based on a linear combination of $e(t)$ and $\dot{e}(t)$, where $\omega_l$ determines the weight of each. Apparently, in closed-loop, $e(t)$ and $\dot{e}(t)$ can be translated to error and its differentiation.\\
\begin{rem}
	\label{rem:stability}
	Following the stability criteria presented in~\cite{dastjerdi2020frequency} for reset elements with shaped error signal, a reset element in CR architecture has the same stability properties as the reset element standing alone. In other words, adding $L(s)$ and $R(s)$ in CR architecture, does not affect the stability properties of the reset element. 
\end{rem}
The output of the proposed architecture is continuous as opposed to  ${\sum{}}_R$ alone. The reason for the continuity of the output of this element is existence of the lag element after the reset element, which makes the discontinuous output of reset element continuous.\\
A known property of the conventional reset elements is existence of high peaks in control input signal, which can result in actuator or amplifier saturation. However, the continuity of the output of the CR element significantly reduces such peaks and consequently saturation problems. 
\section{Proposed Control Loop Architecture}
\label{sec:loop}
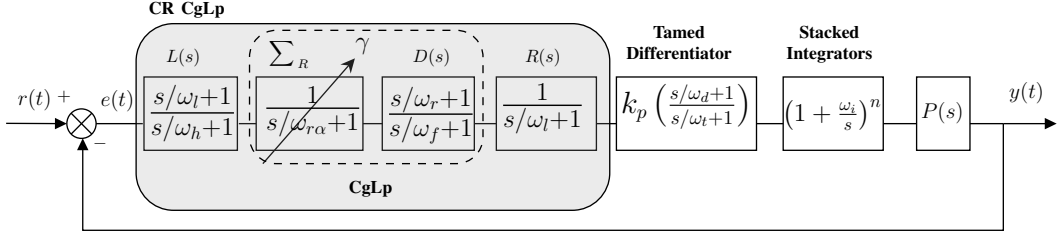
\begin{figure*}[t!]
	\centering
	\vspace{5pt}
	\resizebox{0.8\textwidth}{!}{

		\tikzset{every picture/.style={line width=0.75pt}} 
		
		\begin{tikzpicture}[x=0.75pt,y=0.75pt,yscale=-1,xscale=1]
			
			\draw  [line width=0.75]  (100.9,70) -- (170.9,70) -- (170.9,125) -- (100.9,125) -- cycle ;
			\draw  [line width=0.75]  (185,70) -- (265,70) -- (265,125) -- (185,125) -- cycle ;
			\draw [line width=0.75]    (190,135) -- (257.03,58.26) ;
			\draw [shift={(259,56)}, rotate = 491.13] [fill={rgb, 255:red, 0; green, 0; blue, 0 }  ][line width=0.08]  [draw opacity=0] (10.72,-5.15) -- (0,0) -- (10.72,5.15) -- (7.12,0) -- cycle    ;
			\draw  [line width=0.75]  (680,70) -- (720,70) -- (720,125) -- (680,125) -- cycle ;
			\draw [line width=0.75]    (65,105) -- (100,105) ;
			\draw [line width=0.75]    (745,105) -- (745,185) -- (55,185) -- (55,119) ;
			\draw [shift={(55,116)}, rotate = 450] [fill={rgb, 255:red, 0; green, 0; blue, 0 }  ][line width=0.08]  [draw opacity=0] (8.93,-4.29) -- (0,0) -- (8.93,4.29) -- cycle    ;
			\draw [line width=0.75]    (-2.42,104.78) -- (40.08,104.99) ;
			\draw [shift={(43.08,105)}, rotate = 180.27] [fill={rgb, 255:red, 0; green, 0; blue, 0 }  ][line width=0.08]  [draw opacity=0] (8.93,-4.29) -- (0,0) -- (8.93,4.29) -- cycle    ;
			\draw [line width=0.75]    (745,105) -- (782,105) ;
			\draw [shift={(785,105)}, rotate = 180] [fill={rgb, 255:red, 0; green, 0; blue, 0 }  ][line width=0.08]  [draw opacity=0] (8.93,-4.29) -- (0,0) -- (8.93,4.29) -- cycle    ;
			\draw [line width=0.75]    (172,105) -- (185,105) ;
			\draw [line width=0.75]    (440,105) -- (455,105) ;
			\draw  [line width=0.75]  (455,70) -- (560,70) -- (560,125) -- (455,125) -- cycle ;
			\draw [line width=0.75]    (560,105) -- (580,105) ;
			\draw   (43.08,105) .. controls (43.08,98.89) and (47.98,93.94) .. (54.04,93.94) .. controls (60.09,93.94) and (65,98.89) .. (65,105) .. controls (65,111.11) and (60.09,116.06) .. (54.04,116.06) .. controls (47.98,116.06) and (43.08,111.11) .. (43.08,105) -- cycle ; \draw   (46.29,97.18) -- (61.79,112.82) ; \draw   (61.79,97.18) -- (46.29,112.82) ;
			\draw  [line width=0.75]  (365,70) -- (440,70) -- (440,125) -- (365,125) -- cycle ;
			\draw  [line width=0.75]  (279.5,70) -- (348.5,70) -- (348.5,125) -- (279.5,125) -- cycle ;
			\draw [line width=0.75]    (265,105) -- (280,105) ;
			\draw [line width=0.75]    (349,105) -- (365,105) ;
			\draw [line width=0.75]    (655,105) -- (680,105) ;
			\draw  [fill={rgb, 255:red, 155; green, 155; blue, 155 }  ,fill opacity=0.2 ] (95,58) .. controls (95,42.54) and (107.54,30) .. (123,30) -- (422,30) .. controls (437.46,30) and (450,42.54) .. (450,58) -- (450,142) .. controls (450,157.46) and (437.46,170) .. (422,170) -- (123,170) .. controls (107.54,170) and (95,157.46) .. (95,142) -- cycle ;
			\draw  [dash pattern={on 4.5pt off 4.5pt}] (180,55) .. controls (180,43.95) and (188.95,35) .. (200,35) -- (335,35) .. controls (346.05,35) and (355,43.95) .. (355,55) -- (355,115) .. controls (355,126.05) and (346.05,135) .. (335,135) -- (200,135) .. controls (188.95,135) and (180,126.05) .. (180,115) -- cycle ;
			\draw  [line width=0.75]  (580,70) -- (655,70) -- (655,125) -- (580,125) -- cycle ;
			\draw [line width=0.75]    (720,105) -- (745,105) ;
			
			\draw (19,87) node  [font=\large]  {$r( t)$};
			\draw (762.5,82) node  [font=\large]  {$y( t)$};
			\draw (81,88) node  [font=\large]  {$e( t)$};
			\draw (41,85) node    {$+$};
			\draw (68,120) node    {$-$};
			\draw (189.5,75.57) node [anchor=north west][inner sep=0.75pt]  [font=\huge]   {$\frac{1}{s/\omega _{r\alpha } +1}$};
			\draw (367,73.4) node [anchor=north west][inner sep=0.75pt]  [font=\huge]   {$\frac{1}{s/\omega _{l} +1}$};
			\draw (102.9,74.65) node [anchor=north west][inner sep=0.75pt]  [font=\huge]   {$\frac{s/\omega _{l} +1}{s/\omega _{h} +1}$};
			\draw (258,39) node [anchor=north west][inner sep=0.75pt]  [font=\Large]  {$\gamma $};
			\draw (281.5,75.57) node [anchor=north west][inner sep=0.75pt] [font=\huge]    {$\frac{s/\omega _{r} +1}{s/\omega _{f} +1}$};
			\draw (682,87.4) node [anchor=north west][inner sep=0.75pt]   [font=\large]  {$P( s)$};
			\draw (103,12) node [anchor=north west][inner sep=0.75pt]   [align=left] {\textbf{CR CgLp}};
			\draw (193,40) node [anchor=north west][inner sep=0.75pt]  [font=\normalsize]  {${\displaystyle \sum{}}_{R}$};
			\draw (385,47.4) node [anchor=north west][inner sep=0.75pt]  [font=\normalsize]  {$R( s)$};
			\draw (116,47.4) node [anchor=north west][inner sep=0.75pt]  [font=\normalsize]  {$L( s)$};
			\draw (301,47.4) node [anchor=north west][inner sep=0.75pt]  [font=\normalsize]  {$D( s)$};
			\draw (253,146) node [anchor=north west][inner sep=0.75pt]   [align=left] {\textbf{CgLp}};
			\draw (457,73.4) node [anchor=north west][inner sep=0.75pt]  [font=\LARGE]   {$k_{p}\left(\frac{s/\omega _{d} +1}{s/\omega _{t} +1}\right)$};
			\draw (578,85) node [anchor=north west][inner sep=0.75pt] [font=\Large]    {$\left( 1+\frac{\omega _{i}}{s}\right)^{n}$};
			\draw (461,30) node [anchor=north west][inner sep=0.75pt]   [align=center] {\textbf{Tamed}\\\textbf{Differentiator}};
			\draw (581,30) node [anchor=north west][inner sep=0.75pt]   [align=center] {\textbf{Stacked}\\\textbf{Integrators}};

		\end{tikzpicture}
	}
	\caption{Proposed control loop for motion control with stacked integrators and CR CgLp.}
	\label{fig:closed_block}
\end{figure*}
The proposed closed-loop block diagram for having stacked integrators along with a CR CgLp to compensate the transient overshoot is presented in Fig.~\ref{fig:closed_block}. In order to avoid amplifying the frequencies higher that the bandwidth the differentiator has been tamed~\cite{schmidt2020design}. In the tamed differentiator,
\begin{equation}
	\omega_d=\omega_{c}/a , \quad \omega_t=a\omega_{c}
	\label{eq:wd}
\end{equation}
where $\omega_c$ is the cross-over frequency of the system. Such a tuning allows for the maximum phase lead of the diffrentiator to be achieved at $\omega_c$. The stacked integrators, following the rule of thumb presented in~\cite{schmidt2020design} has a corner frequency of
\begin{equation}
\omega_i=\omega_{c}/10.
\label{eq:wi}
\end{equation}
Such a tuning results in loss of about $5^\circ$ of Phase Margin (PM) for each stacked integrator. Thus, one can expect an increased overshoot and settling time as $n$ increases.  On the other hand, as $n$ increases, the steady-state precision of the system is expected to be increased.\\

\begin{rem}\label{rem:df}
	For large enough $\omega_h$, i.e, $\omega_h \approx \infty$, the CR architecture has the same DF as the ${\sum{}}_R$ alone. 
\end{rem}
It is worth  mentioning that the CgLp element has a constant gain and since according to Remark~\ref{rem:df}, CR architecture will not change the gain and phase behaviour of CgLp, the overall control loop will have a DF gain and phase matching to that of P$\text{I}^\text{n}$D. Thus the control design in open-loop can be done in frequency domain.\\
According to the guidelines presented in~\cite{saikumar2019constant,dastjerdi2021frequency,karbasizadeh2021fractional} for tuning of CgLp elements, the following values can be proposed for CgLp parameters:
\begin{equation}
	\omega_r=\omega_{c},\quad \gamma=0,\quad \alpha =1.1,\quad  \omega_f=20\omega_c.
	\label{eq:cglp_param}
\end{equation}
\begin{figure}[t!]
	\centering
	\includegraphics[width=\columnwidth]{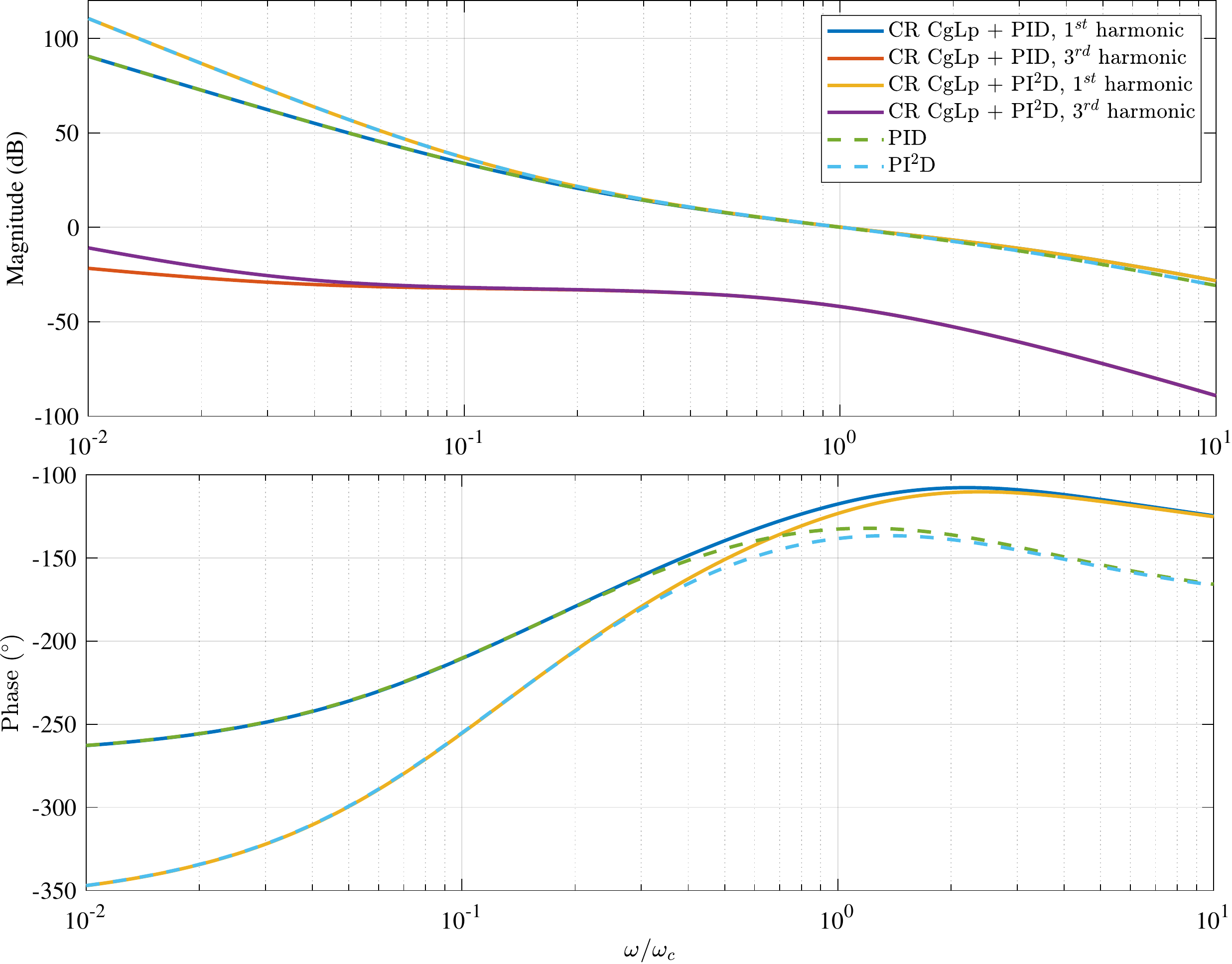}
	\caption{DF and HOSIDF analysis of CR CgLp+PI$^\text{n}$D compared to PI$^\text{n}$D. For the sake of clarity only $n=1,2$ is shown. As a rule of thumb $a=3$.}
	\label{fig:open_bode}
\end{figure}
The open-loop DF and HOSDF analysis of control loop presented in Fig.~\ref{fig:closed_block} for $n=1,2$, tuned according to~\eqref{eq:wd},~\eqref{eq:wi} and~\eqref{eq:cglp_param} is depicted in  Fig.~\ref{fig:open_bode}. Furthermore, in order to observe the effect of CR CgLp, the open-loop Bode plot for the controller in absence of CR CgLp is also depicted in the same plot.\\
Without loss of generality, plant is assumed to be a mass with $P(s)=1/s^2$. As a rule of thumb for Fig.~\ref{fig:open_bode}, $a=3$ for the tamed differentiator. As expected and revealed in Fig.~\ref{fig:open_bode}, the gain of DF for CgLp+PI$^\text{n}$D matches PI$^\text{n}$D, however, a phase lead has been provided. For an added differentiator, both controllers lose around $5^\circ$ of PM. This trend continues as more integrators are stacked but nor shown in the plot for the sake of clarity.
\section{Closed-loop Transient Response}
\label{sec:transient}
\begin{figure}[t!]
	\centering
	\includegraphics[width=\columnwidth]{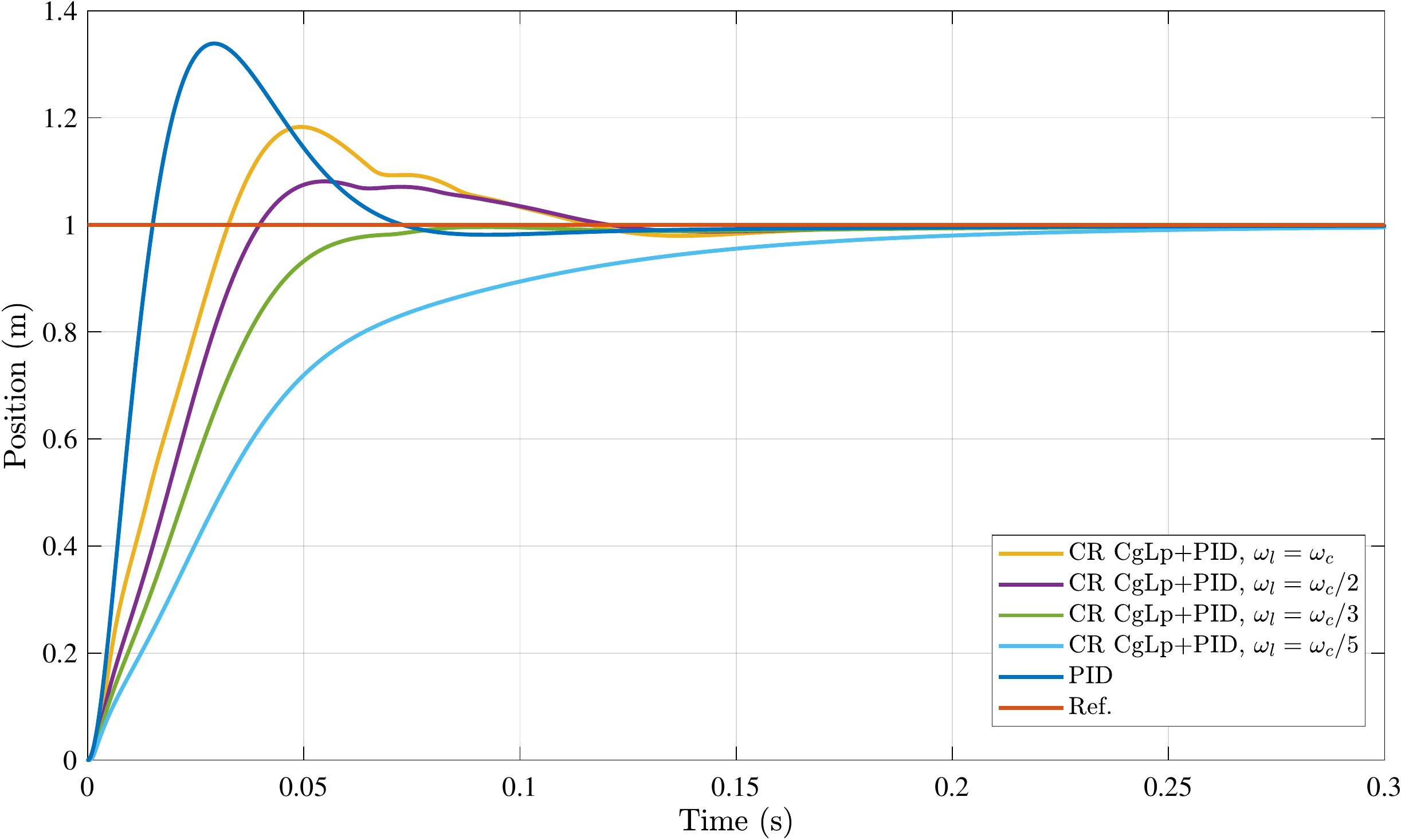}
	\caption{Step response of PID and CR CgLp+PID for different values of $\omega_l$.}
	\label{fig:crcglp_pid_step}
\end{figure}
\begin{figure}[t!]
	\centering
	\includegraphics[width=\columnwidth]{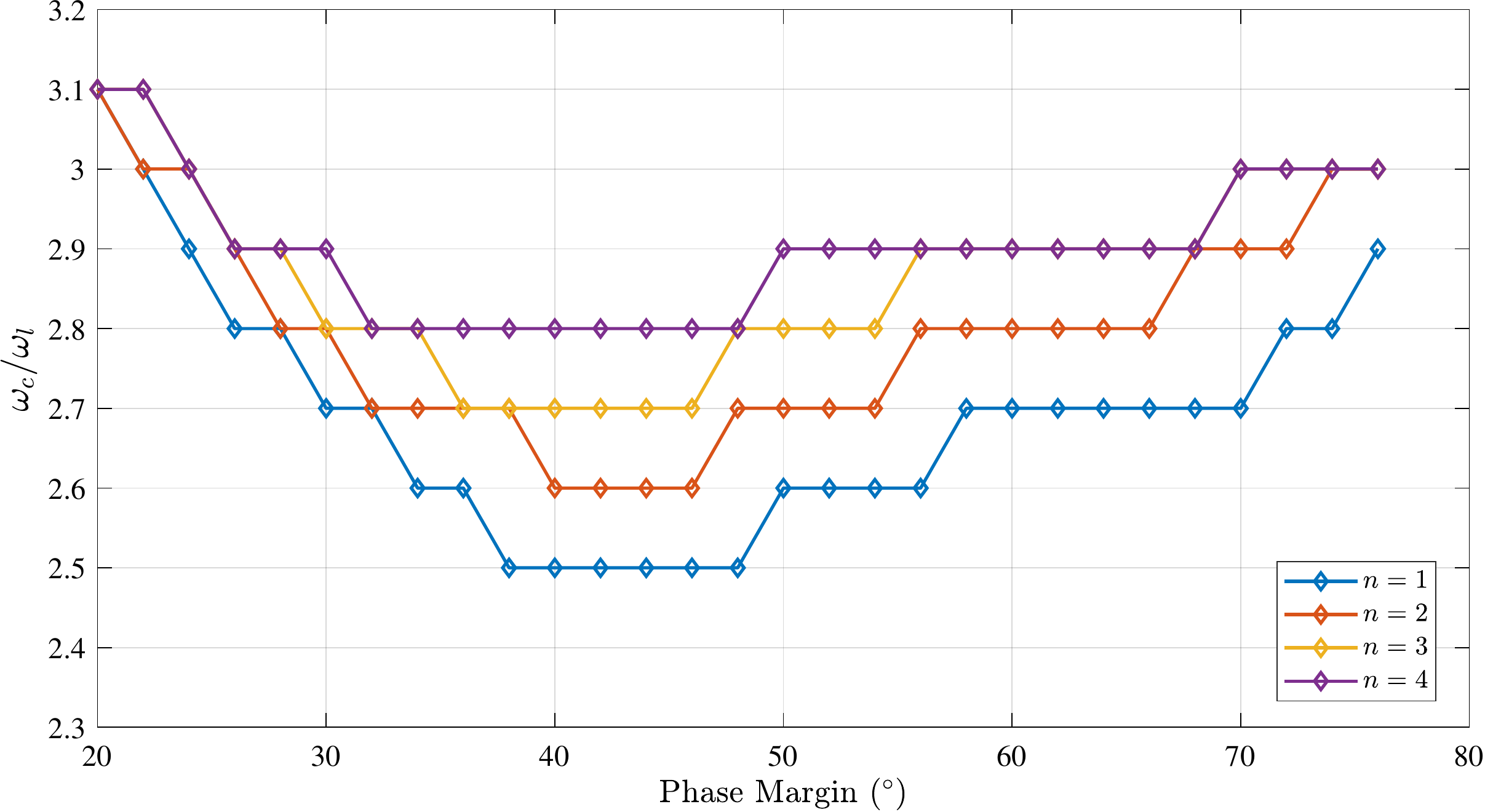}
	\caption{The maximum ratio of $\omega_c/\omega_l$ to achieve a no-overshoot performance for different values of PM and for $n=1,...,4$. Note that $15^\circ$ of PM is provided by CR CgLp. } 
	\label{fig:pm_wl}
\end{figure}
\begin{figure}[t!]
	\centering
	\vspace{5pt}
	\includegraphics[width=\columnwidth]{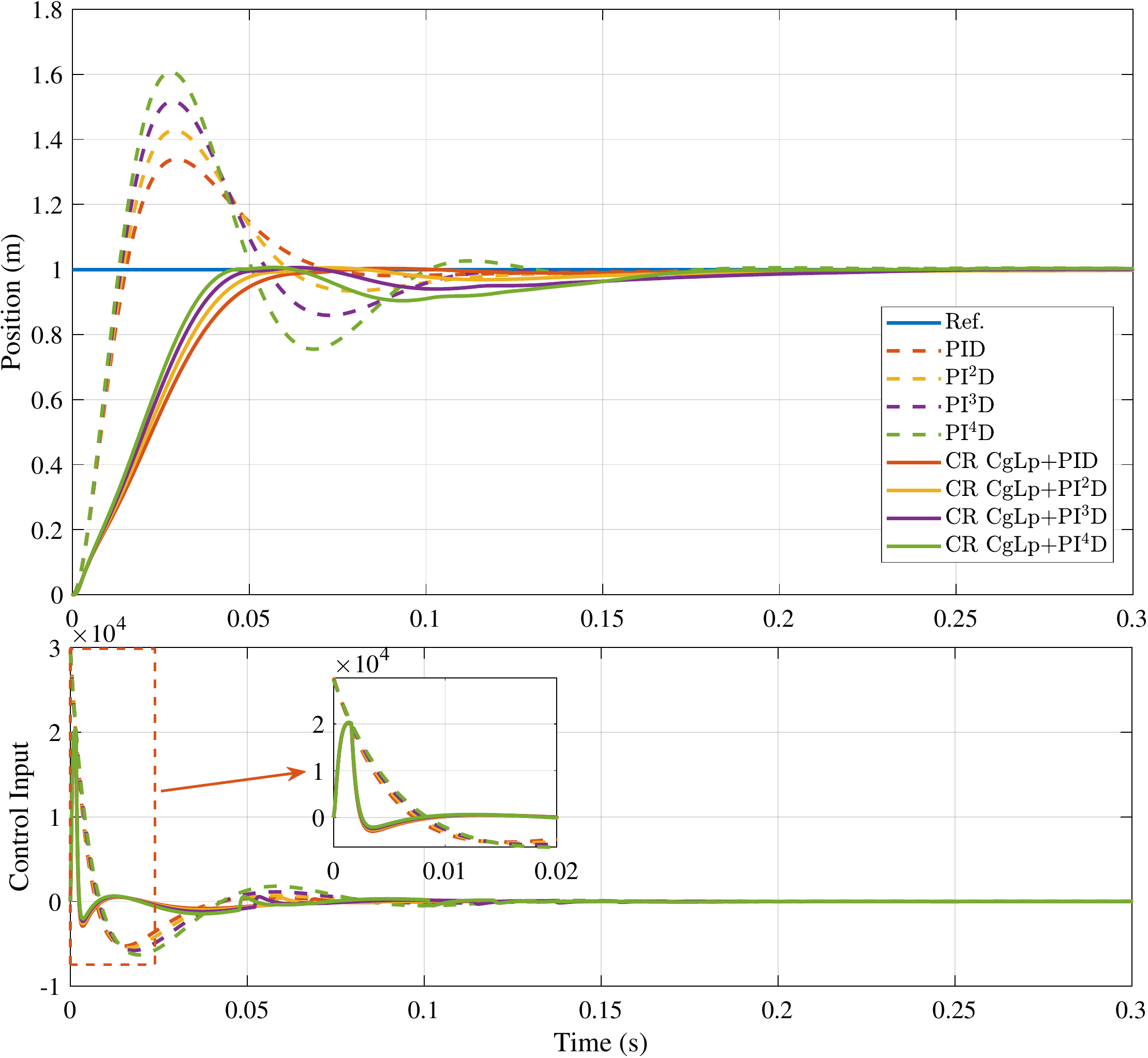}
	\caption{Step response of CR CgLp+PI$^\text{n}$D and PI$^\text{n}$D for $n=1,...,4$. $\omega_{c}=100$ rad/s and $a=3$. The PM for PI$^\text{n}$D is $45-5(n-1)$ and for CR~CgLp+PI$^\text{n}$D is $60-5(n-1)$ degrees. For CR CgLp+PI$^\text{n}$D, $\omega_c/\omega_l=3$.}
	\label{fig:step_n_1_4}
\end{figure}
\begin{figure}[t!]
	\centering
	\vspace{5pt}
	\includegraphics[width=\columnwidth]{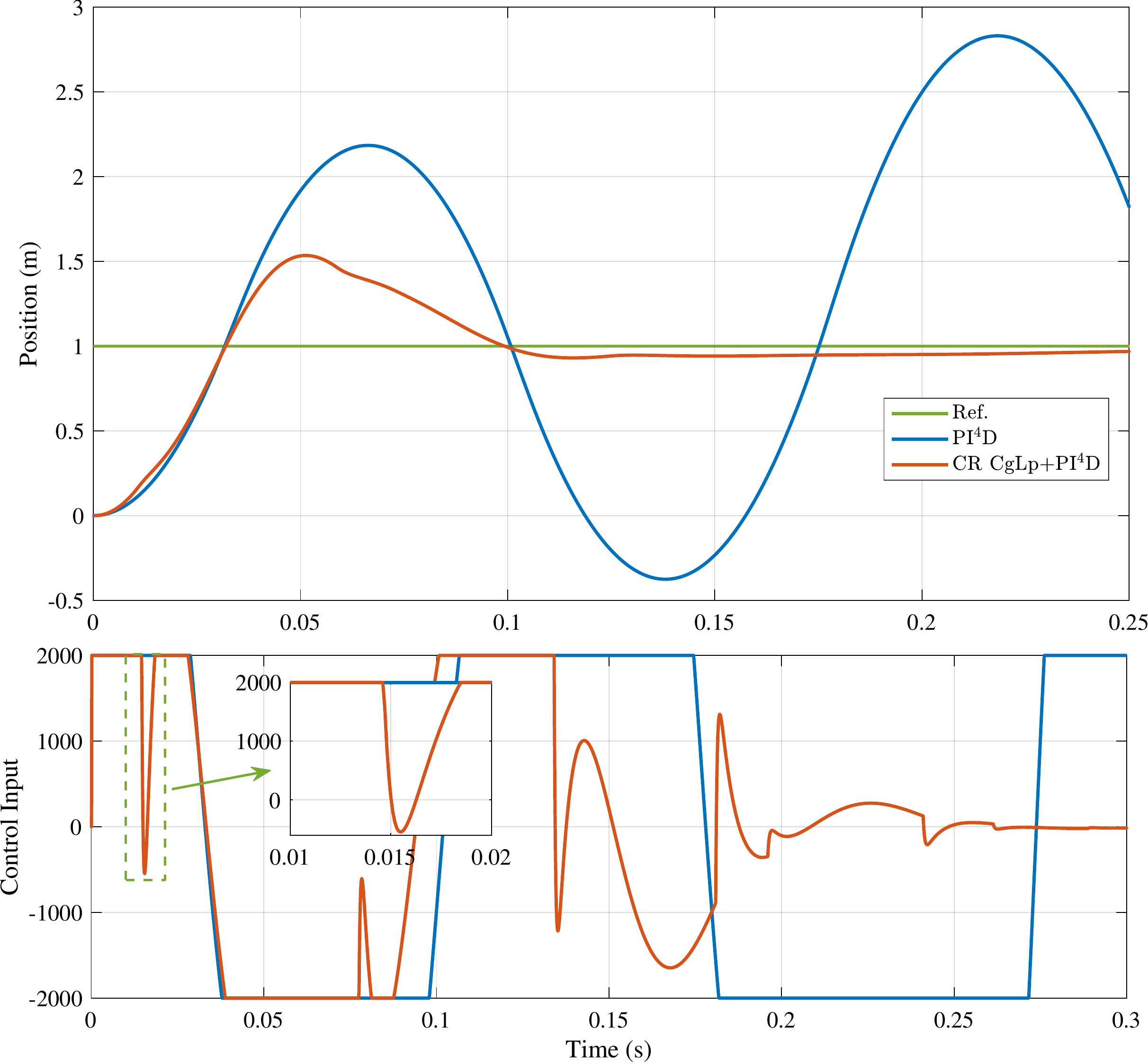}
	\caption{Step response of CR CgLp+PI$^\text{4}$D and PI$^\text{4}$D. Controllers are the same as in Fig.~\ref{fig:step_n_1_4} and the control input signal is saturated at 2000 level.}
	\label{fig:wind-up}
\end{figure}

Because of the phase lead provided by CR CgLp, it is expected that CgLp+PI$^\text{n}$D show a lower overshoot as compared to PI$^\text{n}$D. While this expectation holds, for reset control systems, PM is not the only parameter affecting the transient response. In CR CgLp controllers, $\omega_{l}$ also significantly affects the transient response, since as discussed in~\eqref{eq:reset_law}, the resetting condition in closed-loop changes.\\
The effect of presence of CR CgLp and variation $\omega_{l}$ on the step response of the CR CgLp+PID control system has been simulated in Simulink environment of Matlab and presented Fig.~\ref{fig:crcglp_pid_step}. Without loss of generality, $\omega_{c}$ has been chosen to be 100 rad/s.\\
Fig.~\ref{fig:crcglp_pid_step} reveals that not only the phase lead provided by CgLp improves the transient by reducing the overshoot but also increase of the ratio of $\omega_c/\omega_l$ also reduces the overshoot. It is worth mentioning that changing $\omega_c/\omega_l$, neither affects the gain behaviour of DF nor its phase.\\
Fig.~\ref{fig:crcglp_pid_step} also shows that by increase of $\omega_c/\omega_l$, one can achieve a no-overshoot performance, while such a performance is not achievable by PID no matter how wide the range of differentiation is.  However, too much increase of $\omega_c/\omega_l$ can result in a longer settling time. In order to find the combinations of PM and $\omega_c/\omega_l$ which can result in no-overshoot performance for $n=1,...,4$, a series of simulations has been carried out and the result is presented in Fig.~\ref{fig:pm_wl}. Note that $15^\circ$ of PM is provided by CR CgLp.\\
Fig.~\ref{fig:pm_wl} suggests that even for 4 stacked integrators a no-overshoot performance is achievable. However, for a certain PM, stacking integrators require a higher ration of $\omega_c/\omega_l$ to achieve no-overshoot. As a rule of thumb one can suggest $\omega_c/\omega_l=3$ for PM larger than $25^\circ$.\\
Fig.~\ref{fig:step_n_1_4} shows that step response of CR CgLp+PI$^\text{n}$D and PI$^\text{n}$D for $n=1,...,4$ for a constant differentiation band, i.e., $a$. Figure reveals that while stacking integrators will increase the the overshoot for linear controllers, CR CgLp+PI$^\text{n}$D, maintains a no-overshoot performance. The control input plot also shows another interesting properties of CR CgLp+PI$^\text{n}$D. Although 4 integrators are stacked for CR CgLp+PI$^\text{4}$D, the maximum of its control input is lower than of PID. This indicates that actuator and amplifier saturation problems despite of stacked integrators will not happen for the architecture presented in this paper. \\
Furthermore the wind-up phenomenon is one of the known problems discouraging the usage of multiple integrators in linear domain. In order to study the robustness of CR CgLp+PI$^\text{n}$D and comparison with PI$^\text{n}$D, a saturation block was implemented in simulations to replicate the actuator saturation. The results for $n=4$ which is the one with most vulnerability to wind-up is presented in Fig.~\ref{fig:wind-up}, other controllers are not shown for the sake of clarity. Moreover, the approximate saturation limit which make each of the controllers unstable is presented in Table~\ref{tab:wind-up}. Figure~\ref{fig:wind-up} reveals that although the transient response has been worsened due to saturation, the resetting action prevented the wind-up to create unsuitability for  CR CgLp+PI$^\text{4}$D at this saturation level. Fig.~\ref{fig:wind-up} also shows that continuity of the control input signal is preserved in presence of saturation.
\begin{table}[t]
	\caption{The approximate saturation level which makes the controllers presented in Fig.~\ref{fig:step_n_1_4} unstable. The data is obtained through numerical analysis with Simulink.}
	\label{tab:wind-up}
	\centering
	\begin{tabular}{@{}lllll@{}}
		\toprule
		& $n=1$ & $n=2$ & $n=3$ & $n=4$ \\ \midrule
		PI$^\text{n}$D              & 320   & 750   & 1300  & 2100  \\
		CR~CgLp+PI$^\text{n}$D & 300   & 700   & 1200  & 1900  \\ \bottomrule
	\end{tabular}
\end{table}
\section{Closed-loop Steady-state Performance}
\label{sec:steady}
\begin{figure}[t!]
	\centering
	\includegraphics[width=\columnwidth]{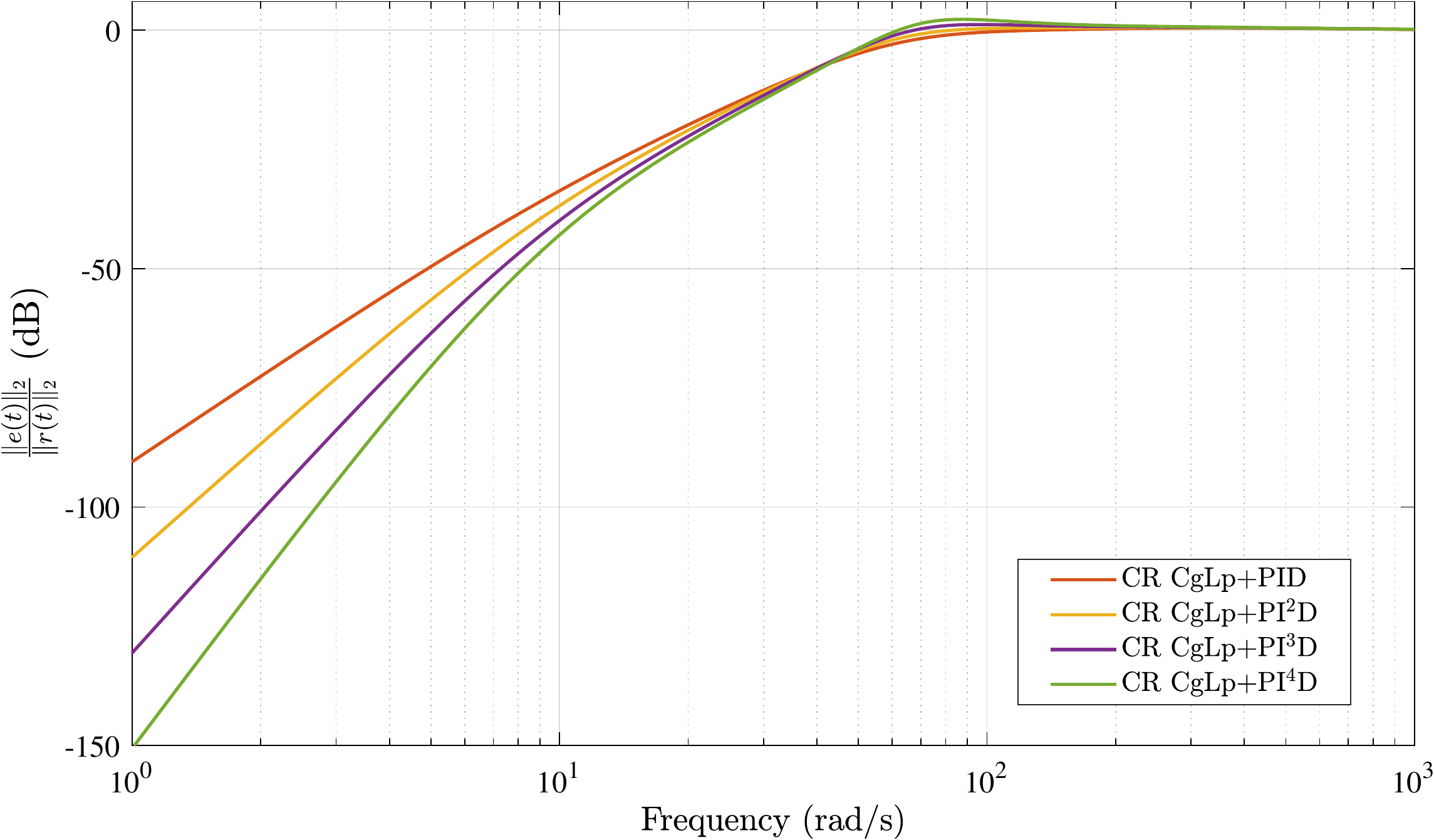}
	\caption{ $\frac{\Vert e(t) \Vert_2}{\Vert r(t) \Vert_2}$ plotted for CR CgLp+PI$^\text{n}$D. The plots closely match to the sensitivity plots calculated based on DF analysis.}
	\label{fig:sensitivity}
\end{figure}
\begin{figure}[!t]
	\centering
	\begin{subfigure}{\columnwidth}
		\includegraphics[width=\textwidth]{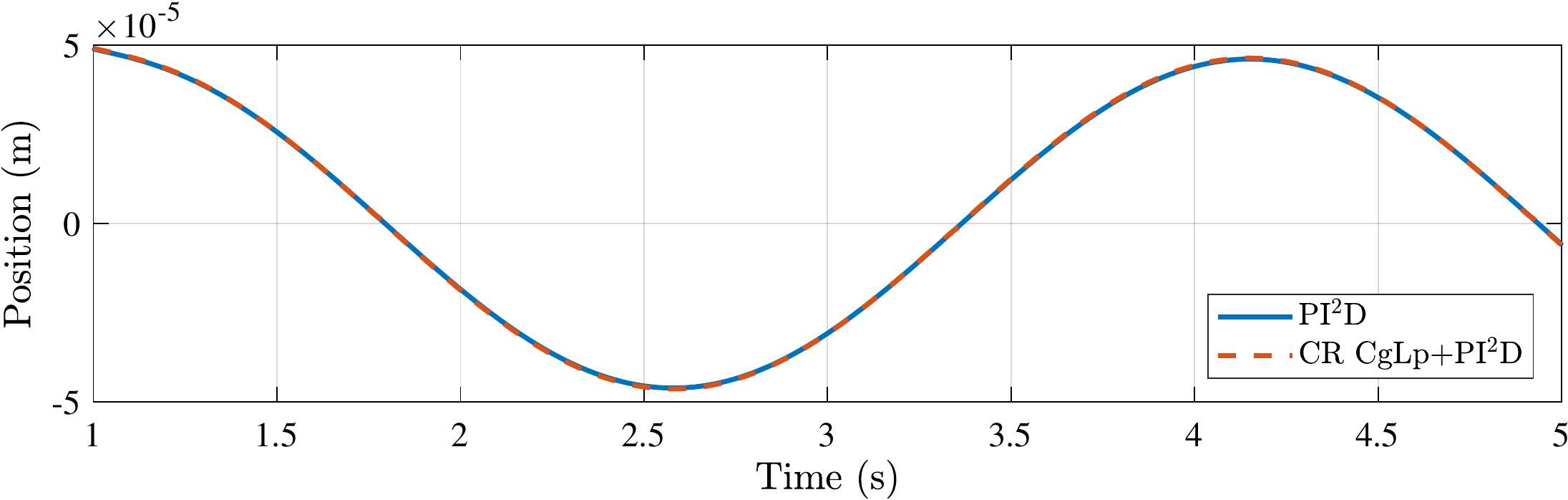}%
		\caption{Steady-state error of the  PI$^\text{2}$D and CR CgLp+PI$^\text{2}$D for a sinusoidal input of $\sin(2t)$.}
		\label{fig:sin_2_n_2}
	\end{subfigure}\\
	\begin{subfigure}{\columnwidth}
		\includegraphics[width=\textwidth]{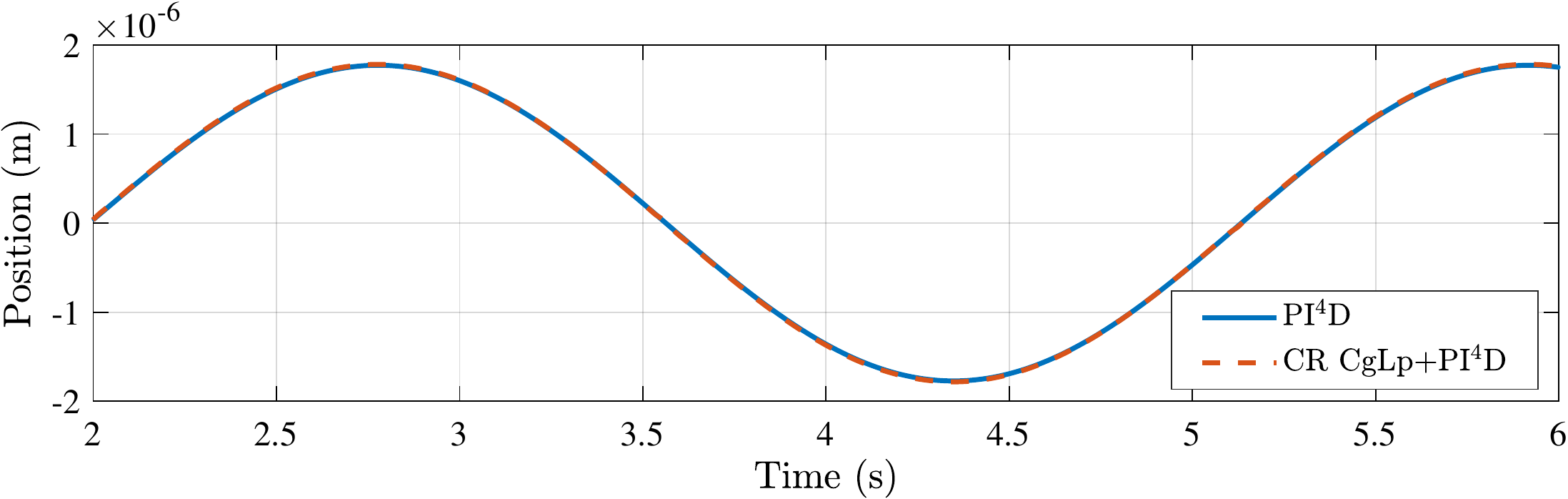}%
		\caption{Steady-state error of the  PI$^\text{4}$D and CR CgLp+PI$^\text{4}$D for a sinusoidal input of $\sin(2t)$. Both plots completely match and are on top of each other.}
		\label{fig:sin_2_n_4}
	\end{subfigure}

	\caption{Steady-state error of the  PI$^\text{n}$D and CR~CgLp+PI$^\text{n}$D for $n=2,4$ to a sinusoidal input of $\sin(2t)$.}
	\label{fig:sin_2}
\end{figure} 
\begin{figure}[!t]
	\centering
	\begin{subfigure}{\columnwidth}
		\includegraphics[width=\textwidth]{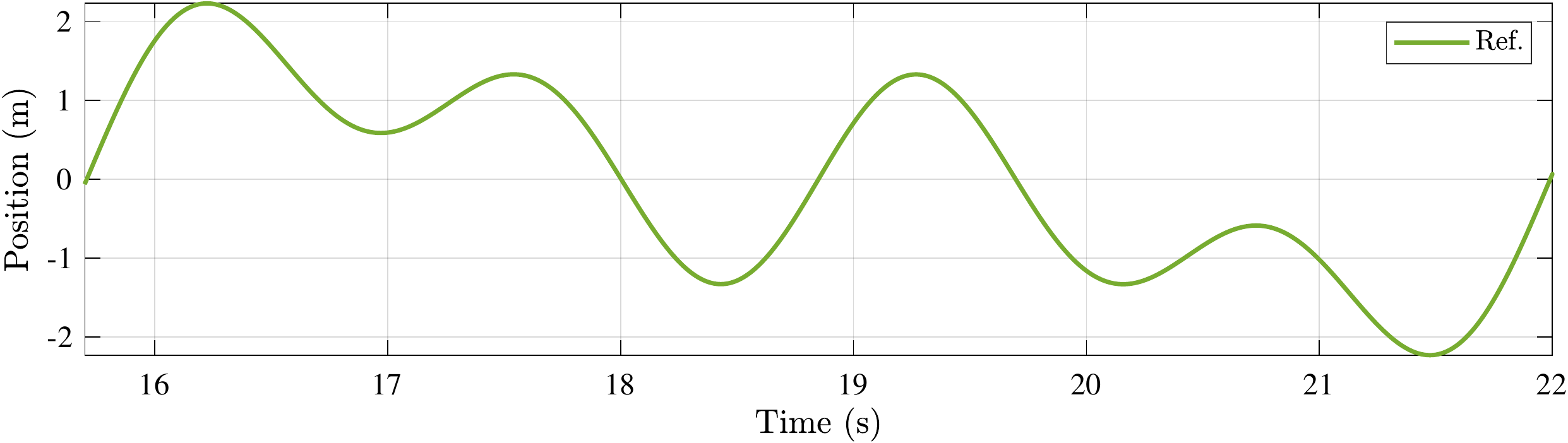}%
		\caption{Multi-sinusoidal reference for tracking comparison of the controllers. $r(t)=\sin(t)+ \sin(2t)+\sin(4t)$.}
		\label{fig:multi_sin_ref}
	\end{subfigure}\\
	\begin{subfigure}{\columnwidth}
		\includegraphics[width=\textwidth]{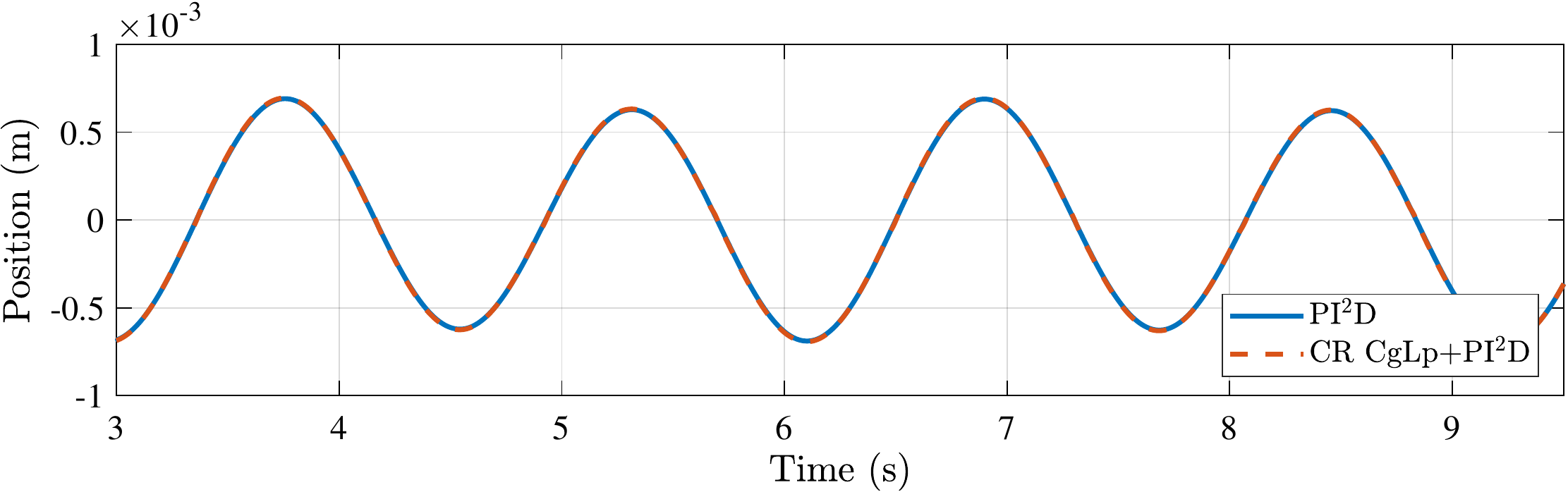}%
		\caption{Steady-state error of the  PI$^\text{2}$D and CR CgLp+PI$^\text{2}$D for a multi-sinusoidal input of~Fig.~\ref{fig:multi_sin_ref}.}
		\label{fig:multi_sin_n_2}
	\end{subfigure}
	\begin{subfigure}{\columnwidth}
		\includegraphics[width=\textwidth]{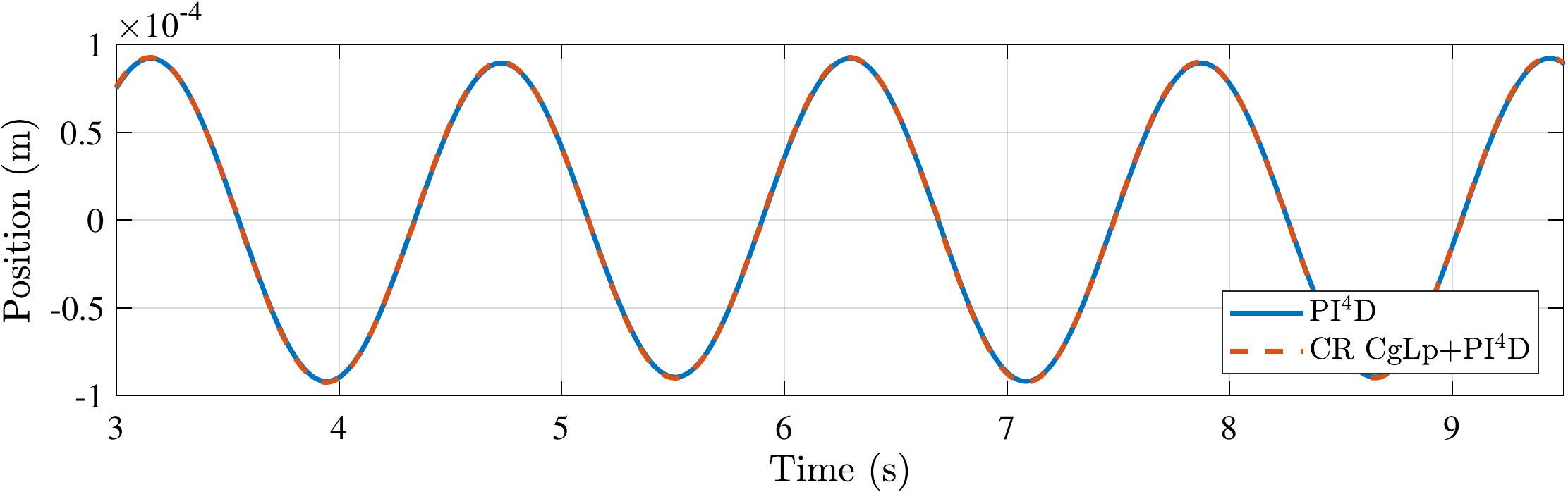}%
		\caption{Steady-state error of the  PI$^\text{4}$D and CR CgLp+PI$^\text{4}$D for a multi-sinusoidal input of~Fig.~\ref{fig:multi_sin_ref}.}
		\label{fig:multi_sin_n_4}
	\end{subfigure}
	
	\caption{Steady-state tracking performance comparison of the  PI$^\text{n}$D and CR~CgLp+PI$^\text{n}$D for $n=2,4$ for a multi-sinusoidal input.}
	\label{fig:multi_sin}
\end{figure} 
The main motivation for stacking integrators is to increase the gain at lower frequencies and consequently decrease the error due to sinusoidal tracking and disturbance resection of the control system. In order to verify the improvement in steady-state performance, one may refer to the sensitivity plot of the control system. However, since reset control systems are nonlinear and sensitivity plot should be approximated for them, one may find sensitivity plot calculated based on DF approximation not accurate~\cite{saikumar2021loop}. In order to more accurately calculate the sensitivity plot, a series of simulations had been carried out for tracking of sinusoidal waves with different frequencies and the $\frac{\Vert e(t) \Vert_2}{\Vert r(t) \Vert_2}$ has been plotted for CR CgLp+PI$^\text{n}$D controllers in  Fig.~\ref{fig:step_n_1_4}. The results are plotted in Fig.~\ref{fig:sensitivity}. The simulation results closely match the sensitivity plots calculated based on DF analysis. Thus, the sensitivity results are not plotted for the sake of clarity.
Fig.~\ref{fig:sensitivity} reveals that the stacked integrators are perfectly reducing the gain at lower frequencies as it would be expected for linear control systems. Thus, the stacked integrators are expected to improve the steady-state precision in terms of tracking and disturbance rejection. \\
In order to better illustrate the steady-state performance of the CR~CgLp+PI$^\text{n}$D controllers, the steady-state error of controllers to sinusoidal input of $\sin(2t)$ is simulated and depicted in Fig.~\ref{fig:sin_2}. The results show that the stacked integrators are performing as expected.\\
Furthermore, the tracking performance of CR~CgLp+PI$^\text{n}$D controllers for a multi-sinusoidal input,
\begin{equation}
	r(t)=\sin(t)+ \sin(2t)+\sin(4t)
\end{equation}
has been compared to that of PI$^\text{n}$D and the steady-state error is depicted in Fig.~\ref{fig:multi_sin}. The comparison verifies that for multi-sinusoidal input as well as sinusoidal ones, the stacked integrators for CR~CgLp+PI$^\text{n}$D, performs as expected.

\section{CONCLUSIONS}

This paper presented a new architecture for a known reset element called CgLp. The new architecture adds a linear lead element before and a linear lag element after the reset element. It was shown that the reset law will be changed due to this change. Furthermore, it was shown that this change significantly improves the transient response of the control system by especially decreasing overshoot. It was shown that the main limitation on linear control systems for stacking multiple investigators, i.e., excessive overshoot, can be solved by adding the proposed reset element to existing linear control loop. A numerical study was done to show the effect of tuning parameters on the transient performance of the proposed reset element in controlling a mass plant and it was shown that even for 4 stacked integrators a no-overshoot performance can be achieved. Moreover, it was shown that the maximum of control input signal for the new architecture is lower than the linear controllers for a similar step input and the new architecture is more robust to unstability which may arise from wind-up phenomenon. The steady-state performance analysis of the proposed control system showed that the main objective of the stacked integrators, i.e., the reduction of sensitivity function gain at lower frequencies is achieved even with presence of the proposed reset element.\\
Practical implementation of the proposed reset element in presence of noise and for more general motion plants such as mass-spring-damper systems is the ongoing work of this research. 
\addtolength{\textheight}{0cm}   





\bibliographystyle{IEEEtran}

\bibliography{ref}

\end{document}